# Self-Attentive Multi-Layer Aggregation with Feature Recalibration and Normalization for End-to-End Speaker Verification System

*Soonshin Seo, Ji-Hwan Kim\**

Dept. of Computer Science and Engineering, Sogang University, Seoul, South Korea
{ssseo,kimjihwan}@sogang.ac.kr

## Abstract

One of the most important parts of an end-to-end speaker verification system is the speaker embedding generation. In our previous paper, we reported that shortcut connections-based multi-layer aggregation improves the representational power of the speaker embedding. However, the number of model parameters is relatively large and the unspecified variations increase in the multi-layer aggregation. Therefore, we propose a self-attentive multi-layer aggregation with feature recalibration and normalization for end-to-end speaker verification system. To reduce the number of model parameters, the ResNet, which scaled channel width and layer depth, is used as a baseline. To control the variability in the training, a self-attention mechanism is applied to perform the multi-layer aggregation with dropout regularizations and batch normalizations. Then, a feature recalibration layer is applied to the aggregated feature using fully-connected layers and nonlinear activation functions. Deep length normalization is also used on a recalibrated feature in the end-to-end training process. Experimental results using the VoxCeleb1 evaluation dataset showed that the performance of the proposed methods was comparable to that of state-of-the-art models (equal error rate of 4.95% and 2.86%, using the VoxCeleb1 and VoxCeleb2 training datasets, respectively).

## 1. Introduction

In the speaker verification field, deep neural networks (DNNs) have been used as speaker embedding extractors. Generally, a speaker embedding-based speaker verification system executes the following process [1–4]:

- First, the classification-based speaker model is trained.
- Second, the speaker embedding is extracted by using the output value of the inner layer of the speaker model.
- Third, the similarity between the embedding of the enrolled speaker and test speaker is computed.
- Fourth, the acceptance or rejection is determined by a pre-decision threshold.

Also, back-end methods, e.g., probabilistic linear discriminant analysis or length normalization, can be used [5–7].

Since the advances in computational power and deep learning techniques, an end-to-end training can demonstrate competitive performance [8–11]. Here, the 'end-to-end' does not refer to a complete end-to-end system, e.g., [12–14], in which a verification result is output when a speech input is given. Herein, it only means that the speaker model training process. Specifically, it is a single-pass training without no additional strategies or back-end methods after extracting the speaker embedding [8, 10].

The most important part of the end-to-end speaker verification system is the speaker embedding generation [10]. A speaker embedding is a high-dimensional feature vector that contains speaker information. An ideal speaker embedding maximizes inter-class variations and minimizes intra-class variations [4, 11, 15]. The component that directly affects the speaker embedding generation is the encoding layer. The encoding layer takes a frame-level feature and converts it into a compact utterance-level feature. It also converts variable-length features to fixed length features.

Most encoding layers are based on a pooling method, e.g. temporal average pooling (TAP) [7, 14, 16], global average pooling (GAP) [10, 15], and statistical pooling (SP) [3, 11, 13, 17, 18]. In particular, self-attentive pooling (SAP) was improved performance by focusing on the frames for a more discriminative utterance-level feature [7, 19, 20]. These pooling layers provide compressed speaker information by rescaling the input size. These are mainly used with convolutional neural networks (CNNs) [7, 10, 11, 14–17, 20]. Therefore, the speaker embedding is extracted by using the output value of the last pooling layer in a CNN-based speaker model.

Furthermore, to improve the representational power of the speaker embedding, residual learning derived from ResNet [21] and squeeze-and-excitation (SE) blocks [22] were adapted for the speaker models [7, 10, 11, 15, 16, 20, 23]. Residual learning maintains input information through mappings between layers called 'shortcut connections.' A large-scaled CNN using shortcut connections can avoid gradient degradation. The SE block consists of a squeeze operation, which condenses all of the information on the features, and an excitation operation, which scales the importance of each features. Therefore, the channel-wise feature response can be adjusted without significantly increasing the model complexity in the training.

The main limitation of previous encoding layers is that the model uses only the output feature of the last pooling layer as input. In other words, it uses only one frame-level feature when constructing a speaker embedding. Therefore, similar to [11, 24], our previous study presented a shortcut connections-based multi-layer aggregation to improve the speaker representations when calculating the weight at the encoding layer [10]. Specifically, the frame-level features is extracted from between each residual layer in ResNet. Then, these frame-level features are fed into the input of the encoding layer using shortcut connections. As a result, a high-dimensional speaker embedding is generated.

---
\* Corresponding author

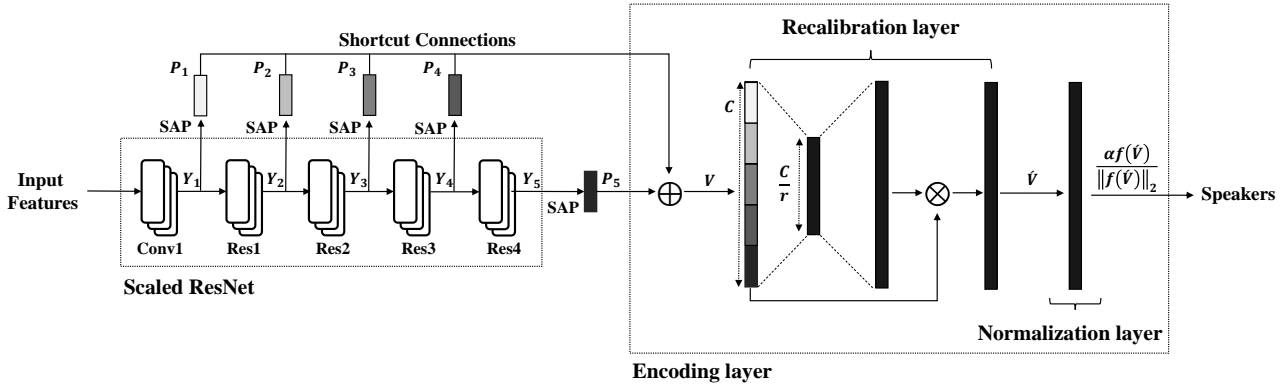

Figure 1: *Network architecture overview: Self-attentive multi-layer aggregation with a feature recalibration layer and a deep length normalization layer. (We extract a speaker embedding after the normalization layer on each utterances.)*

However, our previous study has the limitations. First, the model parameter size is relatively large and the model generates a high-dimensional speaker embeddings (1024-dimensions, about 15 million model parameters). These lead to inefficient training and requires a sufficiently large amount of data for training. Second, the multi-layer aggregation approach increases not only the speaker information, but also the intrinsic and extrinsic variation factors, e.g., emotion, noise, and reverberation. Some of these unspecified factors increase the variability while generating a speaker embedding.

Given that, we propose a self-attentive multi-layer aggregation with feature recalibration and normalization for end-to-end speaker verification system, as shown in Figure 1. We present an improved version of our previous work as described in the following steps:

- First, a ResNet, which scaled channel width and layer depth, is used as a baseline. The scaled ResNet has fewer parameters than the standard ResNet [21].
- Second, a self-attention mechanism is applied to perform the multi-layer aggregation with dropout regularizations and batch normalizations [25]. It helps to construct a more discriminative utterance-level feature, while considering the frame-level features of each layer.
- Third, a feature recalibration layer is applied to the aggregated feature. The channel-wise dependencies are trained using fully-connected layers and nonlinear activation functions.
- Fourth, deep length normalization [8] is also used for a recalibrated feature in the end-to-end training process.

The paper is organized as follows. Section 2 describes a baseline system using shortcut connections-based multi-layer aggregation. Section 3 introduces the proposed self-attentive multi-layer aggregation with feature recalibration and normalization. Section 4 discusses our experiments and conclusions are drawn in Section 5.

## 2. Baseline System: Shortcut Connections-based Multi-Layer Aggregation

### 2.1. Prior system

In our previous study [10], we proposed a shortcut connections-based multi-layer aggregation with ResNet-18. The main difference from standard ResNet-18 [21] is how the speaker embedding is aggregated. The multi-layer aggregation uses not only the output feature of the last residual layer but also the output features of all previous residual layer. These features are concatenated into one feature through shortcut connections. The concatenated feature is fed into several fully connected layers to construct a high-dimensional speaker embedding. Our prior system improved the performance in a simple method; but, model parameters were too large.

Table 1: *Architecture of scaled ResNet-34 using multi-layer aggregation as a baseline (D: input dimension, L: input length, N: number of speakers, SE: speaker embedding)*

| Layer | Output size | Channels | Blocks | Encoding |
|---|---|---|---|---|
| conv1 | $D \times L$ | 32 | - | - |
| pool1 | $1 \times 32$ | - | - | GAP |
| res1 | $D \times L$ | 32 | 3 | - |
| pool2 | $1 \times 32$ | - | - | GAP |
| res2 | $D/2 \times L/2$ | 64 | 4 | - |
| pool3 | $1 \times 64$ | - | - | GAP |
| res3 | $D/4 \times L/4$ | 128 | 6 | - |
| pool4 | $1 \times 128$ | - | - | GAP |
| res4 | $D/8 \times L/8$ | 256 | 3 | - |
| pool5 | $1 \times 256$ | - | - | GAP |
| concat | $1 \times 512$ | - | - | **SE** |
| output | $512 \times N$ | - | - | - |

### 2.2. Modifications

The prior system is modified considering the scaling factors, e.g., layer depth, channel width, and input resolution, for efficient learning in the CNN [26]. First, we use a high-dimensional log-Mel filterbanks with data augmentation for the input resolution. Second, the channel width is reduced and the layer depth is expanded because ResNet can improve the performance without significantly increasing the parameters when the layer depth is increased.

Consequently, the scaled ResNet-34 is constructed as shown in Table 1. The scaled ResNet-34 is composed of three, four, six, and three residual blocks. It is reduced the number of channels by half, compared to the standard ResNet-34 [21]. Also, shortcut connections-based multi-layer aggregation is added to the model using GAP encoding method. The output features of each GAP are concatenated and fed into the output layer. Then, a high-dimensional speaker embedding is generated from a penultimate layer in networks. As a result, the scaled ResNet-34 has only about 6 million model parameters compared to the 12 million of the standard ResNet-18 and 22 million of the standard ResNet-34.

## 3. Self-Attentive Multi-Layer Aggregation with Feature Recalibration and Normalization

### 3.1. Model architecture

As shown in Figure 1 and Table 2, the proposed network mainly consists of a scaled ResNet and an encoding layer. Frame-level features are trained in the scaled ResNet and utterance-level features are trained in the encoding layer.

Table 2: *Architecture of proposed scaled ResNet-34 model using self-attentive multi-layer aggregation with feature recalibration (FR) and deep length normalization (DLN) layers (D: input dimension, L: input length, N: number of speakers, $P$ : output features of pooling layers, $V$ : output features of concatenation layer, $\acute{V}$: output features of FR layer, SE: speaker embedding)*

| Layer | Output size | Channels | Blocks | Encoding |
|---|---|---|---|---|
| conv1 | $D \times L$ | 32 | - | - |
| pool1 | $1 \times 32$ | - | - | SAP ($P_1$) |
| res1 | $D \times L$ | 32 | 3 | - |
| pool2 | $1 \times 32$ | - | - | SAP ($P_2$) |
| res2 | $D/2 \times L/2$ | 64 | 4 | - |
| pool3 | $1 \times 64$ | - | - | SAP ($P_3$) |
| res3 | $D/4 \times L/4$ | 128 | 6 | - |
| rool4 | $1 \times 128$ | - | - | SAP ($P_4$) |
| res4 | $D/8 \times L/8$ | 256 | 3 | - |
| pool5 | $1 \times 256$ | - | - | SAP ($P_5$) |
| concat | $1 \times 512$ | - | - | $V$ |
| FR | $1 \times 512$ | | | $\acute{V}$ |
| DLN | $1 \times 512$ | - | - | SE |
| output | $512 \times N$ | - | - | - |

**Scaled ResNet.** In the scaled ResNet, given an input feature $X = [x_1, x_2, ..., x_l, ..., x_L]$ of length $L$ ( $x_l \in \mathbb{R}^d$ ), output features $P_i = [p_1, p_2, ..., p_c, ..., p_C]$ ( $p_c \in \mathbb{R}^1$ ) from each residual layer of scaled ResNet are generated using SAP.

Here, the length $C_i$ is determined by the number of the last channel in the $i^{th}$ residual layer. Then, the generated output features are concatenated in order into one feature $V$ ([+] indicates concatenation)

$$V = P_1 \,[+]\, P_2 \,[+]\, P_3 \,[+]\, P_4 \,[+]\, P_5. \quad (1)$$

The concatenated feature $V = [v_1, v_2, ..., v_c, ..., v_C]$ (length $C = C_1 + C_2 + C_3 + C_4 + C_5$, $v_c \in \mathbb{R}^1$) is a set of frame-level features and is used as the input of the encoding layer.

**Encoding layer.** The encoding layer consists of a feature recalibration layer and a deep length normalization layer. In the feature recalibration layer, the concatenated feature $V$ is recalibrated by fully connected layers and nonlinear activations. As a result, a recalibrated feature $\acute{V} = [\acute{v}_1, \acute{v}_2, ..., \acute{v}_c, ..., \acute{v}_C]$ ( $\acute{v}_c \in \mathbb{R}^1$) is generated.

Then, the recalibrated feature is normalized according to the length of the input $\acute{V}$ in the deep length normalization layer. The normalized feature is used to a speaker embedding and is fed into the output layer for discriminating speaker classes.

### 3.2. Self-attentive multi-layer aggregation

As shown in Figure 1, the SAP is applied to each residual layer using shortcut connections. Given an output feature of the first convolution layer or the $i^{th}$ residual layers after conducting an average pooling, $Y_i = [y_1, y_2, ..., y_n, ..., y_N]$ of length $N$ ($y_n \in \mathbb{R}^c$) is obtained. The number of dimension $c$ is determined by the number of channels.

Then, the average feature is fed into a fully-connected hidden layer to obtain $H_i = [h_1, h_2, ..., h_n, ..., h_N]$ using a hyperbolic tangent activation function. Given $h_n \in \mathbb{R}^c$ and a learnable context vector $u \in \mathbb{R}^c$, the attention weight $w_n$ is measured by training the similarity between $h_n$ and $u$ with a softmax normalization as

$$w_n = \frac{exp(h_n^T \cdot u)}{\sum_{n=1}^{N} exp(h_n^T \cdot u)}. \quad (2)$$

Then, the embedding $e \in \mathbb{R}^d$ is generated using the weighted sum of the normalized attention weights $w_n$ and $h_n$ as

$$e = \sum_{n=1}^{N} h_n w_n. \quad (3)$$

In this manner, the SAP output feature $P_i = [p_1, p_2, ..., p_c, ..., p_C]$ ($p_c \in \mathbb{R}^1$) is generated. This process helps to generate a more discriminative feature while paying attention to the frame-level features of each layer. Moreover, dropout regularization and batch normalization are used in feature $P_i$. Then, the generated features are concatenated into one feature $V$ as in equation (1).

### 3.3. Feature recalibration

After the self-attentive multi-layer aggregation, concatenated feature $V$ is given to the feature recalibration layer as input. The feature recalibration layer aims to train the correlations between each channel of the concatenated feature; this is inspired by [22].

Given an input feature $V = [v_1, v_2, ..., v_c, ..., v_C]$ ($v_c \in \mathbb{R}^1$, where $C$ is the sum of all channels), the feature channels are recalibrated using two fully-connected layers and nonlinear activations as follows:

$$\acute{V} = f_{FR}(V, W) = \sigma(W_2 \delta(W_1 V)). \quad (4)$$

Here, $\delta$ refers to the leaky ReLU activation and $\sigma$ refers to the sigmoid activation. $W_1$ is the front fully-connected layer, $W_1 \in \mathbb{R}^{c \times \frac{c}{r}}$, and $W_2$ is the back fully-connected layer, $W_2 \in \mathbb{R}^{\frac{c}{r} \times c}$. According to the reduction ratio $r$, a dimensional transformation is performed between the two fully-connected layers, such as a bottleneck structure. Also, the channel-wise multiplication is performed. Then, the rescaled channels are multiplied by the input feature $V$. As a result, an output feature $\acute{V} = [\acute{v}_1, \acute{v}_2, ..., \acute{v}_c, ..., \acute{v}_C]$ ( $\acute{v}_c \in \mathbb{R}^1$ ) is generated. It is recalibrated according to the importance of the channels.

### 3.4. Deep length normalization

Like [8], deep length normalization is applied to proposed model. The L2-constraint is applied to the length axis of recalibrated feature $\acute{V}$ with scale parameter $\alpha$ as

$$\frac{\alpha f_{DLN}(\acute{V})}{\|f_{DLN}(\acute{V})\|_2} = f_{DLN}(\acute{V}). \quad (5)$$

Then, the normalized $\acute{V}$ is fed into the output layer for speaker classification. This feature is used as a speaker embedding.

## 4. Experiments

### 4.1. Datasets

**Training.** In our experiments, we used the VoxCeleb1 [27] and VoxCeleb2 [16] training datasets, which were collected in real environments. These are large-scale text-independent speaker verification datasets, consisting of more than 100 thousand and 1 million utterances with 1211 and 5994 speakers, respectively.

**Evaluation.** We used the VoxCeleb1 evaluation dataset, which includes 40 speakers and 37,220 pairs of the official test protocol. This is an open-set test, which evaluates all speaker pairs that are not seen in the training dataset.

### 4.2. Experimental setup

**Input features.** We used a 64-dimensional log Mel-filterbanks with a 25 ms frame length and 10 ms frame shift, which were mean-variance normalized over a sliding window of up to 3 s. For each training step, 12 s interval was extracted from each utterance using cropping or padding.

**Preprocessing.** In the training, a SpecAugment method was used to perform time and frequency masking on input features [28].

**Hyper-parameters.** We used a cross entropy-based softmax loss function. Also, we used the stochastic gradient descent optimizer with a momentum of 0.9, a weight decay of 0.0001, and an initial learning rate of 0.1, reduced by a 0.1 scale factor on the plateau. All experiments were trained for 200 epochs with a 96 mini-batch size. The scaling parameter $\alpha$ was initialized to a value of 10 and the reduction ratio $r$ was initialized to a value of 8.

**Evaluation metrics.** From the trained model, we generated a 512-dimensional speaker embedding in each utterance. We used a standard cosine similarity for computing the speaker pair. We used the equal error rate (EER, %) and minimum detection cost function (minDCF) as evaluation metrics.

### 4.3. Experimental results

To evaluate the proposed methods, we first tested the baseline using different encoding methods. We then compared our proposed method with state-of-the-art encoding methods.

Table 3: *Experimental results for modifying the baseline construction, using the VoxCeleb1 training dataset (Dim: speaker embedding dimension).*

| Model | Encoding method | Dim | EER | minDCF |
|---|---|---|---|---|
| Scaled ResNet-34 | GAP | 256 | 6.85 | 0.3389 |
| | SAP | 256 | 6.68 | 0.3402 |
| | MLA-SAP | 512 | 5.42 | 0.3025 |
| | MLA-SAP-FR | 512 | 5.07 | 0.2932 |
| | MLA-SAP -FR-DLN | 512 | 4.95 | 0.2902 |

**First comparison.** Table 3 presents the results of modifying the baseline described in Table 1. We experimented with basic encoding layers, e.g., GAP and SAP. Then, we combined the proposed methods one by one to the baseline, e.g., self-attentive multi-layer aggregation (MLA-SAP), feature recalibration (FR), and deep length normalization (DLN).

**Second comparison.** Table 4 shows a comparison of our proposed methods with state-of-the-art encoding methods. Here, we focused on encoding methods using a ResNet-based model and the softmax loss function in the end-to-end training process. Various encoding methods were compared, e.g., TAP [7, 16], learnable dictionary encoding (LDE) [7], SAP [7], GAP [15], NetVLAD [4], and GhostVLAD [4]. The experimental results showed that our proposed methods was superior to the state-of-the-art methods (EER of 4.95% and 2.86%, using the VoxCeleb1 and VoxCeleb2 training datasets, respectively).

Table 4: *Experimental results comparing our proposed methods with state-of-the-art encoding methods using the VoxCeleb1 and VoxCeleb2 training datasets*

| Model | Encoding method | Dim | EER |
|---|---|---|---|
| ResNet-34 [7] * | TAP | 128 | 5.48 |
| ResNet-34 [7] * | LDE | 128 | 5.21 |
| ResNet-34 [7] * | SAP | 128 | 5.51 |
| ResNet-34 [15] * | GAP | 256 | 5.39 |
| Scaled ResNet-34 (proposed) * | MLA-SAP -FR-DLN | 512 | **4.95** |
| ResNet-34 [16] | TAP | 512 | 5.04 |
| ResNet-50 [16] | TAP | 512 | 4.19 |
| Thin-ResNet-34 [4] | NetVLAD | 512 | 3.57 |
| Thin-ResNet-34 [4] | GhostVLAD | 512 | 3.22 |
| Scaled ResNet-34 (proposed) | MLA-SAP -FR-DLN | 512 | **2.86** |

* These models used the VoxCeleb1 training datasets.

## 5. Conclusions

In previous multi-layer aggregation methods for end-to-end speaker verification, the number of model parameters had relatively large and unspecified variations increased in the training. Therefore, we proposed a self-attentive multi-layer aggregation with feature recalibration and normalization for end-to-end speaker verification system. First, the ResNet, which scaled channel width and layer depth, was used as a baseline. Second, a self-attentive multi-layer aggregation was applied when training the frame-level features of each residual layer in the scaled ResNet. Third and fourth, the feature recalibration layer and deep length normalization were applied to train the utterance-level feature in the encoding layer. The experimental results using the VoxCeleb1 evaluation dataset showed that the proposed method achieved a performance comparable to state-of-the-art models.

## 6. Acknowledgements

This work was supported by Institute for Information & communications Technology Promotion (IITP) grant funded by the Korea government (MSIT) (No.2017-0-01772, Development of QA systems for Video Story Understanding to pass the Video Turing Test)